\documentclass[prl,twocolumn,epsf,psfig]{revtex4}
\usepackage{graphicx}

\begin{document}

\title{Breathing mode frequencies of a rotating Fermi gas in the BCS-BEC crossover region}

\author{Theja N. De Silva}
\affiliation{Department of Physics, Applied Physics and Astronomy,
The State University of New York at Binghamton, Binghamton, New York
13902, USA.}
\begin{abstract}
We study the breathing mode frequencies of a rotating Fermi gas
trapped in a harmonic plus radial quartic potential. We find that as
the radial anharmonicity increases, the lowest order radial mode
frequency increases while the next lowest order radial mode
frequency decreases. Then at a critical anharmonicity, these two
modes merge and beyond this merge the cloud is unstable against the
oscillations. The critical anharmonicity depends on both rotational
frequency and the chemical potential. As a result of the large
chemical potential in the BCS regime, even with a weak anharmonicity
the lowest order mode frequency increases with decreasing the
attractive interaction. For large enough anharmonicities in the weak
coupling BCS limit, we find that the excitation of the breathing
mode frequencies make the atomic cloud unstable.
\end{abstract}

\maketitle

\section{I. Introduction}
The rapid progress of ultra-cold atomic gas experiments provides
unique opportunities for well controlled studies of quantum many
body physics. For the case of Fermi atomic systems, the possibility
of controlling the s-wave scattering length ($a$) between two
different spin components allows to control the interaction by using
a magnetically tuned Feshbach resonance~\cite{fb}. This unique
capability allows one to investigate the cross-over between the
weakly interacting BCS regime (the regime where $a \rightarrow 0^-$)
and Bose-Einstein condensate of dimers (the regime where $a
\rightarrow 0^+$)~\cite{co}. These two regimes meet in strongly
interacting limit where the scattering length is divergent and at
this unitarity limit, the physics is expected to be
universal~\cite{uni}.

The appearance of quantized vortices of a quantum fluid under
rotation offers direct evidence of superfluidity. For example, the
observation of quantized circulation in a rotating superfluid
$^4$He~\cite{vortexHe} and the observation of vortex lattice in a
rotating Fermi gas of $^6$Li~\cite{vortexMIT} are two classic
demonstrations of phase coherence in a superfluid. These are analog
to the vortex lattice in type-II superconductors in the presence of
a magnetic filed. These vortices melt as the magnetic field
increases and then the superconductors turn into normal at
sufficiently large magnetic fields. For the case of rapidly rotating
Fermi gasses, the force due to the trapping frequency almost
balances the centrifugal force and superfluid cloud spreads in the
plane perpendicular to the rotation axis. At the limit of very large
rotation, the theory predicts that the atomic system enters into the
fractional quantum Hall regime~\cite{ho, fqhe}. However, the
fractional quantum Hall window is expected to be very small and
inversely proportional to the number of atoms in the trap. A
possible way of stabilizing the fractional quantum Hall regime is to
add a positive quartic trapping potential.

In this paper, we study the collective breathing mode frequencies of
a rotating Fermi gas in the presence of a quartic trapping potential
by using a hydrodynamic approach. A negative, but small quartic term
is always present with the Gaussian optical potentials in current
experimental setups while added positive quartic term ensures the
stability of the fast rotating regime. Thermodynamic properties of a
Bose gas confined in harmonic plus quartic potential trap can be
found in ref.~\cite{quarticBose}. As breathing mode frequencies are
very sensitive to the equation of state, these dynamical quantities
can be used as tests for various theories. The breathing mode
frequencies have been measured for non-rotating Fermi systems in the
BCS-BEC region~\cite{modesEX1,modesEX2}. In most of the parameter
regions, experimental results agree well with the hydrodynamic
approaches, variational approaches and sum rule
approaches~\cite{modesTH}. However, the measured finite temperature
axial and radial breathing mode frequencies show a striking increase
in the intermediate BCS regime~\cite{modesEX1,modesEX2}, in contrast
to zero temperature theoretical calculations in a harmonic trap.
Furthermore, experimentalists were unable to measure the breathing
mode frequencies in the weak coupling BCS limit. The deviation of
the experimental data from theoretical results and the lack of
experimental data in the weak coupling BCS limit were believed to be
due the large Landau damping when the superfluid energy gap is much
smaller than the collective oscillation energies. With inclusion of
a positive quartic term in the trapping potential, we find somewhat
similar deviation of the breathing mode frequencies in the
intermediate regions of BCS regime. We find that the breathing mode
frequencies deviate significantly from the modes frequencies
calculated in a harmonic trap and the atomic cloud is unstable
against the breathing mode oscillations at larger chemical
potentials. As the chemical potential is larger in the weak coupling
BCS limit, excitation of breathing mode frequencies make the atomic
cloud unstable. It should be noted that the Gaussian optical trap
potential provides a negative quartic term in the trapping potential
in experimental setups. We investigate the effect of negative
quartic term and find that the breathing mode frequencies tend to
decrease in the entire BCS-BEC crossover region.

This paper is organized as follows. In section II, we present the
derivation of breathing mode frequencies using a hydrodynamic
approach. In section III, we present our results together with a
discussion. Finally in section IV, we draw our conclusions.

\section{II. Formalism}

We consider a rotating Fermi atomic system trapped in a harmonic
plus radial quartic potential in the BCS-BEC crossover region. The
trapping potential is

\begin{equation}\label{trap}
V_{ex}(r,z) = \frac{1}{2}M\omega_r^2r^2 + \frac{1}{2}M\omega_z^2z^2
+ \frac{K}{4}r^4
\end{equation}

\noindent where $M$ is the atom mass, $\omega_i$'s are the harmonic
trapping frequencies, and $r^2 = x^2+y^2$. The Fermi atomic system
rotates about the $z$-axis at frequency $\Omega$.

We restrict ourselves to the case of large number of vortices in the
system where the wavelength of the oscillation frequencies is much
larger than the inter-vortex distance. This condition always
satisfies as the typical wavelength of the lowest mode oscillations
is of the order of the system size. Further, we assume that the
vortices are uniformly distributed in the superfluid so that we do
not have to consider microscopic details of the single vortices.
These assumptions are valid at the limit of large rotations where
the atomic cloud spreads in the plane perpendicular to the rotation.
Within this diffused vorticity approximation~\cite{vorticity},
diffuse vorticity is given by $\mathbf{\nabla} \times \mathbf{v} =
2\mathbf{\Omega}$, where superfluid velocity is given by $
\mathbf{v} = (\hbar/2M)\nabla \theta$. The local superfluid density
$n$ and the local phase $\theta$ are related through the wave
function $\psi = \sqrt{n}e^{i\theta}$. The uniform vortex density is
given by $n_v = 2M\Omega /\hbar$.

Assuming local equilibrium, we start with the continuity and Euler
equations of rotational hydrodynamics,

\begin{equation}\label{continuity}
\frac{\partial n}{\partial t}=-\mathbf{\nabla} \cdot [n(r)
\mathbf{v}]
\end{equation}
and
\begin{eqnarray}\label{euler}
M\frac{\partial \mathbf{v}}{\partial
t}=-\mathbf{\nabla}[\frac{1}{2}M\mathbf{v}^2+V_{ex}(r,z) -
\frac{1}{2}\Omega^2r^2 + \mu(n)] \nonumber \\
+ 2M \mathbf{v} \times
\mathbf{\Omega} + M\mathbf{v }\times \mathbf{\nabla} \times
\mathbf{v}
\end{eqnarray}

This hydrodynamic description is valid as long as the collisional
relaxation time $\tau$ is much smaller than the inverse of the
oscillation frequencies; $\omega \tau << 1$. The equation of state
enters through the density dependent local chemical potential
$\mu(n)$. We fix the local chemical potential by introducing the
equation of state in the form of $\mu(n) \propto n^{\gamma}$.  As we
will discuss in the next subsection, the polytropic index $\gamma$
is calculated by the method proposed by Manini and
Salasnich~\cite{manini} in the entire BCS-BEC crossover region.
Linearizing the density $n$ and the superfluid velocity $\mathbf{v}$
around their equilibrium values as $n=n_0(r) + \delta n$,
$\mathbf{v}=\mathbf{v}_0 + \delta \mathbf{v}$ and $\mu(n) = \mu(n_0)
+ \delta \mu$ with $\delta \mu = (\partial \mu/\partial n)|_{n=n_0}
\delta n$, we obtain the linearized version of the hydrodynamic
equations.

\begin{equation}\label{linear1}
\frac{\partial \delta n}{\partial t}=-\mathbf{\nabla }\cdot [n_0(r)
\delta \mathbf{v}]
\end{equation}
and
\begin{equation}\label{linear2}
M\frac{\partial \delta \mathbf{v}}{\partial t}=-\mathbf{\nabla}
\delta \mu+ 2M \delta \mathbf{v} \times \mathbf{\Omega}
\end{equation}

Starting from these two linearized equations, collective breathing
mode frequencies have been calculated in ref.\cite{gosh} and
ref.\cite{ant} for a harmonic trap. As the authors have used two
different ansatz for the velocity fluctuation, they produce two
different results for the breathing mode frequencies. In this paper
we closely follow the approach adopted in ref.\cite{ant},
generalizing the theory to an anharmonic trap. The ansatz used in
ref.\cite{ant} ensures the conservation of angular momentum
properly. In order to solve the linearized equations for the
breathing mode frequencies, we take the equilibrium density in the
local density approximation as $n_0(r) \propto
[\mu_0-(1/2)M(\omega_r^2-\Omega_0^2)r^2-(1/2)M\omega_z^2z^2-(K/4)r^4]^{1/\gamma}$
and use following variational ansatz for the density fluctuations
and velocity fluctuations.

\begin{eqnarray}\label{ansatz1}
\delta \mathbf{v}=\{\delta \mathbf{\Omega_1} \times
\mathbf{r}+\delta \mathbf{\Omega_2} \times r^2
\mathbf{r}+\mathbf{\nabla}[\alpha_{\perp}r^2+\alpha_zz^2+\beta
r^4]\}\nonumber \\ \times \exp[-i\omega t]
\end{eqnarray}
and
\begin{equation}\label{ansatz2}
\delta n=n_0^{1-\gamma}\{a_0+a_{\perp}r^2+a_zz^2+br^4\}\exp[-i\omega
t]
\end{equation}
The first two terms $\delta \mathbf{\Omega_1}$ and $\delta
\mathbf{\Omega_2}$ in Eq.~(\ref{ansatz1}) are parallel to the axis
of rotation and guarantee that angular momentum is conserved during
the oscillations. Substituting these two ansatz into
Eq.~(\ref{linear1}) and Eq.~(\ref{linear2}), we derive four linear
equations for the variational parameters $a_0$, $a_{\perp}$, $a_z$,
and $b$. These linear equations yield three non-zero solutions for
the breathing mode frequencies $\omega_m \equiv \omega/\omega_r$ as
roots of the following equation.

\begin{equation}\label{sol}
A+B\omega_m^2+C\omega_m^4+\omega_m^6 = 0
\end{equation}

with, $A \equiv -(1-\zeta^2)^2 \delta^2
[64d\gamma(\gamma+1)+48\gamma^2+56 \gamma+16]-(1-\zeta^2) \zeta^2
\delta^2(32 \gamma^2+104 \gamma+48)-16 \zeta^4 \delta^2(\gamma+2)$,
$B \equiv (1-\zeta^2)^2[8(\gamma+1)(2\gamma+1)+16d\gamma
(\gamma+2)]+(1-\zeta^2) [\delta^2 (8\gamma^2+26
\gamma+12)+\zeta^2(40\gamma+24)]+8\zeta^2\delta^2
(\gamma+2)+16\zeta^4$ and $C \equiv
(1-\zeta^2)(2-10\gamma)-\delta^2(\gamma+2)-8$. The constants, $\zeta
\equiv \Omega/\omega_r$, $\delta \equiv \omega_z/\omega_r$, and $d
\equiv [1/(1-\zeta^2)^2] (K \hbar/M^2 \omega_r^3)(\mu_0/\hbar
\omega_r)$ are a set of dimensionless parameters. The three
solutions of Eq.~(\ref{sol}) are the lowest order axial breathing
mode frequency $\omega_1$ and the lowest and next lowest order
radial breathing mode frequencies $\omega_2$ and $\omega_3$.

\subsection{The Effective polytropic index and the chemical potential in the BCS-BEC crossover region}

We use the proposal made by Manini and Salasnich~\cite{manini} to
calculate the effective polytropic index $\gamma$ and the chemical
potential $\mu$ in the BCS-BEC crossover region. In the weak
coupling BCS limit ($a \longrightarrow 0^-$) and the unitarity limit
($a \longrightarrow \infty$), the polytropic index is $\gamma =
2/3$. In the deep BEC limit ($a \longrightarrow 0^+$), the
polytropic index is $\gamma = 1$. In the BCS-BEC crossover regime,
the scattering length's dependence on $\gamma$ is given
by~\cite{manini}

\begin{equation}\label{gamma}
\gamma = \frac{2/3-2y\epsilon^{\prime}(y)/5+y^2\epsilon^{\prime
\prime}(y)/15}{\epsilon(y)-y\epsilon^{\prime}(y)/5}
\end{equation}

\noindent with the parameter $y = 1/(k_fa)$ is the interaction
parameter with $k_f$ being the Fermi wave vector. The function
$\epsilon(y)$ is related to the energy per atom given by $E =
(3/5)E_f\epsilon(y)$, where $E_f = \hbar^2k_f^2/2M$ is the Fermi
atomic energy of a non-interacting Fermi system in the trap.
 Above $\epsilon^{\prime}(y) = \partial
\epsilon(y)/\partial y$ and the double prime indicates the second
derivative of the function on its argument. Using the data presented
in reference~\cite{astra}, Manini and Salasnich~\cite{manini} used a
data fitting scheme to derive an analytical form of the function
$\epsilon(y)$ in the entire BCS-BEC region,

\begin{equation}\label{epsilon}
\epsilon(y) = \alpha_1-\alpha_2 \arctan\biggr[\alpha_3 y
\frac{\beta_1+|y|}{\beta_2+|y|}\biggr]
\end{equation}
Two different sets of parameters are proposed for $\alpha_i$'s and
$\beta_i$'s in the BCS regime ($y < 0$) and the BEC regime ($y >
0$). In the BCS regime, the parameters are $\alpha_1 = 0.42$,
$\alpha_2 = 0.3692$, $\alpha_3 = 1.044$, $\beta_1 = 1.4328$, and
$\beta_2 = 0.5523$. In the BEC regime, the parameters are $\alpha_1
= 0.42$, $\alpha_2 = 0.2674$, $\alpha_3 = 5.04$, $\beta_1 = 0.1126$,
and $\beta_2 = 0.4552$. The expression for the chemical potential
$\mu$ is given by~\cite{manini}

\begin{equation}\label{mu}
\mu = E_f[\epsilon(y)-y\epsilon^{\prime}(y)/5]
\end{equation}

We determine the Fermi energy $E_f$ of a non-interacting Fermi
system in a harmonic plus radial quartic potential through the
number equation.

\begin{equation}\label{ef}
N = \frac{1}{15 \delta}\sqrt{\frac{M^2\omega_r^3}{\hbar
|K|}}\biggr(\frac{E_f}{\hbar \omega_r}\biggr)^{5/2}f(E_f)
\end{equation}

\noindent We define the function $f(E_f)$ as
\begin{eqnarray}\label{func}
f(E_f) = \pm 8 \biggr(\pm
1+\frac{(1-\zeta^2)^2}{4}\frac{1}{|\tilde{K}|\tilde{E_f}}\biggr)^{\frac{5}{2}}
\nonumber \\ \mp
\sqrt{\frac{(1-\zeta^2)^2}{4}\frac{1}{|\tilde{K}|\tilde{E_f}}}
\nonumber \\
\times
\biggr[15+5(1-\zeta^2)^2\frac{1}{\tilde{K}\tilde{E_f}}+\frac{1}{2}
(1-\zeta^2)^4\frac{1}{(\tilde{K}\tilde{E_f})^2}\biggr]
\end{eqnarray}

\noindent where the scaled parameters are $\tilde{K} \equiv \hbar
K/(M^2\omega_r^3)$ and $\tilde{E_f} \equiv E_f/(\hbar \omega_r)$.
The upper and lower signs are corresponding to $K >0$ and $K<0$
respectively. For the case of harmonic trap, the Fermi energy is
$E_f=\hbar\omega_r[3N\delta(1-\zeta^2)]^{1/3}$.

\begin{figure}
\includegraphics[width=\columnwidth]{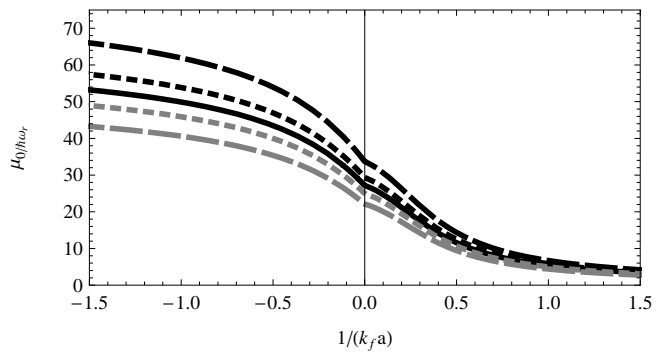}
\caption{Central chemical potential ($\mu_0$) of a non-rotating
Fermi gas as a function of interaction parameter $1/(k_fa)$ for
various values of $\tilde{K}$. From top to bottom $\tilde{K} = 0.05$
(long black dashed line),
 and $0.01$ (short black dashed line), $0$ harmonic trap (black solid line), $-0.005$
(short gray dashed line),
 and $-0.01$ (long gray dashed line). For the calculation, we use $N = 2.0
\times 10^6$ and $\delta = 0.045$.} \label{pmu}
\end{figure}

\section{III. results and discussion}

\begin{figure}
\includegraphics[width=\columnwidth]{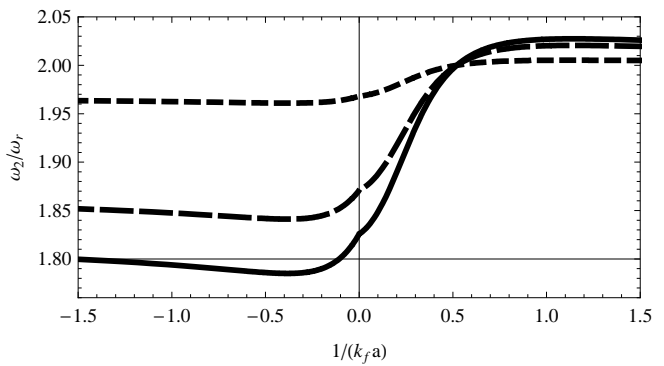}
\caption{The lowest order radial breathing mode frequencies of a
rotating Fermi gas in a harmonic trap with aspect ratio $\delta =
0.045$. The rotation frequencies are $\zeta = 0$ (solid line), $0.5$
(long dashed line) and $0.9$ (short dashed line).} \label{pradial}
\end{figure}

In FIG.~\ref{pmu}, we plot the central chemical potential of a
non-rotating Fermi system calculated from Eq.~(\ref{mu}) as a
function of inverse scattering length for two different
representative values of anharmonicity. We use $N = 2.0 \times 10^6$
number of atoms in the trap with $\delta = 0.045$. Our calculation
shows a kink at unitarity limit due to the discontinuity in the
function $\epsilon (y)$ proposed by Manini and
Salasnich~\cite{manini} (This kink appears in all the calculated
macroscopic quantities).

Solving Eq.~(\ref{sol}) for the case of harmonic potential trap ($K
= 0$) in a rotating ($\zeta \neq 0$) Fermi system, the lowest axial
and radial breathing modes frequencies are given by

\begin{eqnarray}\label{solrot}
\omega_m^2=
\frac{1}{2}\{\gamma(2+\delta^2-2\zeta^2)+2(1+\delta^2+\zeta^2)
\nonumber \\
 \pm \{[\gamma(2+\delta^2-2\zeta^2)+2(1+\delta^2+\zeta^2)]^2 -
 \nonumber \\
8\delta^2[\gamma(-3+\zeta^2)-2(1+\zeta^2)]\}^{1/2}\}.
\end{eqnarray}

For the isotropic trap ($\delta = 1$), at non-interacting limit and
at unitarity limit ($\gamma =2/3$), the mode frequencies are
$\omega_m = 2$ and $\omega_m = \sqrt{2+2\zeta^2/3}$. In the deep BEC
limit ($\gamma = 1$), the mode frequencies are $\omega_m =
\sqrt{(1/2)(7 \pm \sqrt{9-8\zeta^2})}$. For the case of harmonic
potential ($K = 0$) in a non-rotating ($\zeta = 0$) limit,
Eq.~(\ref{solrot}) reduces to

\begin{eqnarray}\label{solnonrot}
\omega_m^2= \frac{1}{2}\{2(1+\delta^2)+\gamma(2+\delta^2) \nonumber
\\ \pm\sqrt{[2(1+\delta^2)+\gamma
(2+\delta^2)]^2-8(2+3\gamma)\delta^2}\}.
\end{eqnarray}

\begin{figure}
\includegraphics[width=\columnwidth]{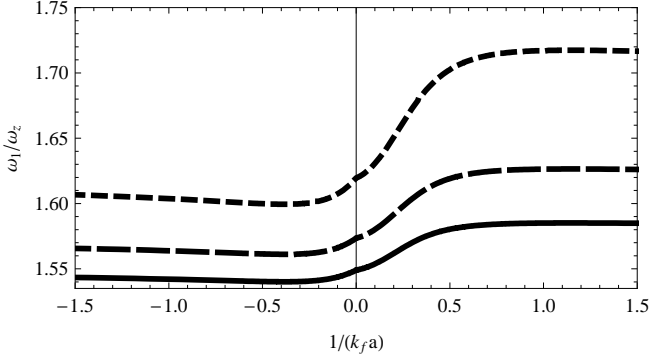}
\caption{The lowest order axial breathing mode frequencies of a
rotating Fermi gas in a harmonic trap with aspect ratio $\delta =
0.045$. The rotation frequencies are $\zeta = 0$ (solid line), $0.5$
(long dashed line) and $0.9$ (short dashed line).} \label{paxial}
\end{figure}

\noindent For the case of highly anisotropic limit ($\delta << 1$),
the two mode frequencies at $\gamma = 2/3$ are $\omega_1/\omega_z =
\sqrt{12/5}$ and $\omega_2/\omega_r = \sqrt{10/3}$ as expected. For
this case, the two mode frequencies at $\gamma = 1$ are
$\omega_1/\omega_z = \sqrt{5/2}$ and $\omega_2/\omega_r = 2$. In the
BCS-BEC crossover region, the lowest order breathing modes
frequencies are calculated from Eq.~(\ref{solnonrot}) by using the
$\gamma$ from Eq.~(\ref{gamma}). The results for several
representative values of $\zeta$ are given in FIG.~\ref{pradial} and
FIG.~\ref{paxial}.

\begin{figure}
\includegraphics[width=\columnwidth]{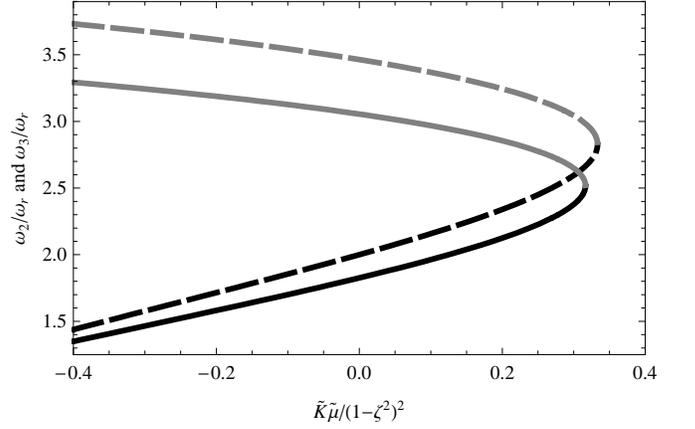}
\caption{The two lowest radial breathing mode frequencies as a
function of $d = [1/(1-\zeta^2)^2] \tilde{K} \tilde{\mu}$. Black
(lowest mode) and gray (second lowest) solid lines are the mode
frequencies for weakly interacting limit and unitarity limit($\gamma
= 2/3$). Black (lowest mode) and gray (second lowest) dashed lines
are the mode frequencies for deep BEC limit($\gamma = 1$). The value
of $\delta = \omega_z/\omega_r = 0.045$. The two modes merge at some
critical $d$ and beyond this critical value, the atomic cloud is
unstable against the breathing mode oscillations.} \label{pstable}
\end{figure}

We solve Eq.~(\ref{sol}) for the breathing mode frequencies for
various values of $K$ in both rotating and non-rotating Fermi
systems. We calculate the central chemical potential $\mu_0$ for
fixed number of atoms $N = 2.0 \times 10^6$ in the trap. We find
that as $d \equiv [1/(1-\zeta^2)^2] (K \hbar/M^2
\omega_r^3)(\mu_0/\hbar \omega_r) = [1/(1-\zeta^2)^2] \tilde{K}
\tilde{\mu}$ increases, the lowest order radial breathing mode
frequency increases while the next lowest order breathing mode
frequency decreases. Then at a critical value of $d =d_c$, these two
modes merge and beyond this critical $d_c$, the atomic cloud is
unstable against the oscillations. FIG.~\ref{pstable} shows the
lowest and the next lowest order radial mode frequencies at $\gamma
=2/3$ (non-interacting limit and unitarity limit)  and $\gamma =1$
(deep BEC limit).

\begin{figure}
\includegraphics[width=\columnwidth]{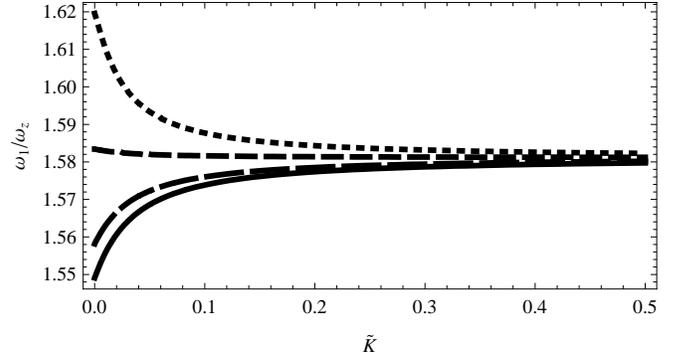}
\caption{Axial breathing mode frequencies as a function of
$\tilde{K}$ for $\zeta = 0$ (black), 0.3 (long dashed), 0.6 (short
dashed) and, 0.9 (dotted) at unitarity. We use the values $\delta =
0.045$ and $N = 2.0 \times 10^6$.} \label{paxialk}
\end{figure}

As evidenced by FIG.~\ref{paxialk}, the lowest order axial breathing
mode frequencies are almost insensitive to the radial
anharmonicities. FIG.~\ref{paxialk} shows the axial breathing mode
frequencies for various values of rotational frequencies as a
function radial anharmonicities at unitarity.

\begin{figure}
\includegraphics[width=\columnwidth]{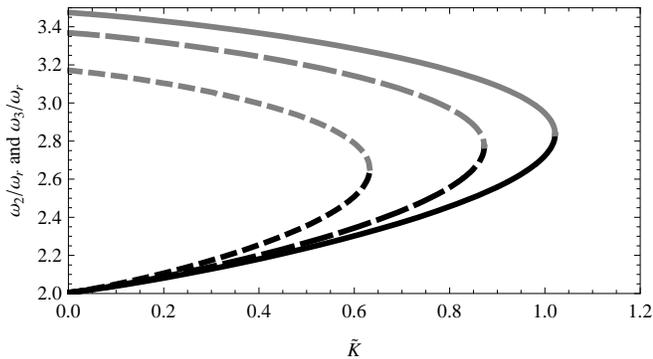}
\caption{Radial breathing mode frequencies at BEC limit ($\gamma =
1$) for $\zeta = 0$ (solid line), $0.3$ (long dashed line) and $0.5$
(short dashed line). The black lines are lowest breathing mode and
the gray lines are second lowest breathing mode in an anharmonic
trap. The value of $\delta = \omega_z/\omega_r = 0.045$ and the atom
number is $N = 2.0 \times 10^6$. } \label{pradanharmbec}
\end{figure}

In the BEC limit where $\gamma = 1$, we calculate the lowest and
next lowest radial breathing mode frequencies as a function of
radial anharmonicity $\tilde{K}$ for three representative values of
rotational frequencies $\zeta$. We fixed the number of atoms to be
$2.0 \times 10^6$ and $\delta = \omega_Z/\omega_r = 0.045$. As shown
in FIG.~\ref{pradanharmbec}, as we increase $\tilde{K}$ the lowest
order radial breathing mode frequency increases, while the next
lowest order radial breathing mode frequency decreases. Further
increase of $\tilde{K}$ merges these two modes and beyond this
merging point the atomic cloud is unstable against the oscillations.

\begin{figure}
\includegraphics[width=\columnwidth]{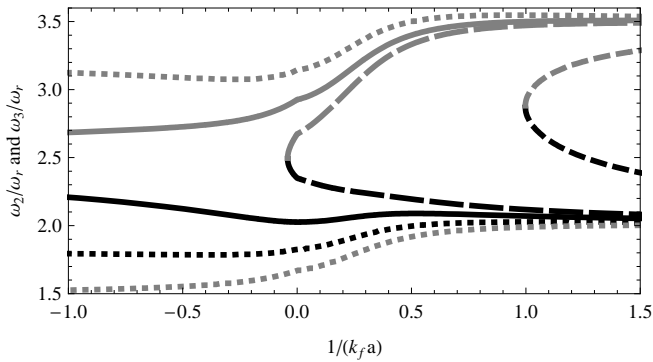}
\caption{The two lowest radial breathing mode frequencies at for
$\tilde{K}= 0.005$ (solid line), $0.01$ (long dashed line) and
$0.05$ (short dashed line). The black lines are lowest breathing
mode and the gray lines are second lowest breathing mode in an
anharmonic trap. The black dotted line is the lowest breathing mode
frequency in a harmonic trap. The gray dotted lines are the
breathing mode frequencies at $\tilde{K}= -0.005$. The value of
$\delta = \omega_z/\omega_r = 0.045$ and the atom number is $N = 2.0
\times 10^6$. } \label{pradanharmuni}
\end{figure}

\begin{figure}
\includegraphics[width=\columnwidth]{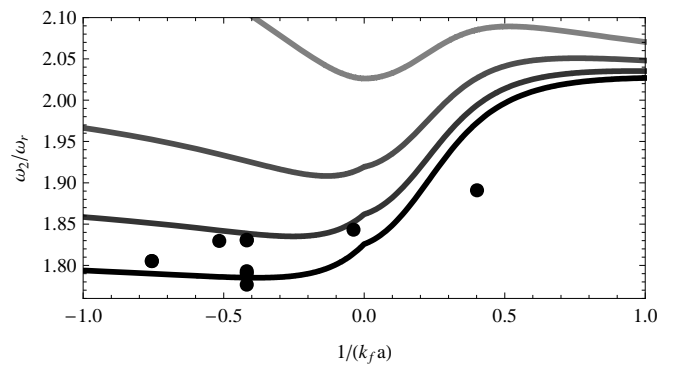}
\caption{The lowest order radial breathing mode frequencies at for
$\tilde{K}= 0.005$, $0.025$ and $0.001$ (Top to bottom). The black
lines is the lowest order breathing mode in a harmonic trap. The
value of $\delta = \omega_z/\omega_r = 0.045$ and the atom number is
$N = 2.0 \times 10^6$. The dots are the experimental data for
comparison~\cite{modesEX2}.} \label{mm}
\end{figure}

The lowest order and next lowest order radial breathing mode
frequencies as a function of the interaction parameter [$1/(k_fa)$]
are shown in FIG.~\ref{pradanharmuni}. In the presence of radial
anharmonicity, the lowest order mode frequency tends to increase in
the BCS regime while the next lowest order breathing mode frequency
tends to decrease. This deviation becomes large as the anharmonicity
increases. As we have discussed before, at larger $K\mu_0$ values
the two lowest order modes merge and the cloud is unstable against
the breathing mode oscillations beyond this point. The data in the
FIG.~\ref{mm} shows the same information as
FIG.~\ref{pradanharmuni}, but we plot only the lowest order radial
breathing mode frequencies for small quartic potentials together
with experimental data from ref.~\cite{modesEX2}.

\section{IV. Conclusions}
We have discussed the breathing mode frequencies of a rotating Fermi
gas trapped in a harmonic plus radial quartic potential. We find
that the radial breathing mode frequencies strongly depend on the
rotation and anharmonicity through parameter $d = [1/(1-\zeta^2)^2]
(K \hbar/M^2 \omega_r^3)(\mu_0/\hbar \omega_r)$. As $d$ increases,
the lowest order radial breathing mode's frequency increases and the
next lowest order mode decreases. Beyond some critical $d_c$, these
two modes merge and the cloud is unstable against the oscillations.

As the chemical potential is large in the intermediate BCS regime,
even with a very weak quartic potential the parameter $d$ is large.
As a result, the lowest order breathing mode frequency increases in
the intermediate BCS regime. Even though the Gaussian optical trap
potential provides a negative anharmonic term in the trapping
potential, this positive anharmonic behavior has been seen in recent
experiments~\cite{modesEX1, modesEX2}. In the weak coupling BCS
limit, the chemical potential is even larger so that we find the
atomic cloud is unstable against the oscillations at large positive
anharmonicities. For negative quartic potentials, the breathing mode
frequencies tend to decrease in the BCS-BEC crossover region.

\section{V. Acknowledgements}
This work was supported by the Binghamton University. We are very
grateful to Kaden Hazzard for very enlightening discussions and
critical comments on the manuscript.


\begin{references}
\bibitem{fb} U. Fano, Phys. Rev. A 124, 1866 (1961); H Feshbach,
Ann. Phys. 5, 357 (1961).
\bibitem{co} C. A. Regal et al., Phys. Rev. Lett. 92, 040403 (2004);
M. W. Zwierlein et al., Phys. Rev. Lett. 92, 120403 (2004); C. Chin
et al., Science 305, 1128 (2004); T. Bourdel et al., Phys. Rev.
Lett. 93, 050401 (2004); J. Kinast et al., Phys. Rev. Lett. 92,
150402 (2004); G. B. Partridge, et al., Phys. Rev. Lett. 95, 020404
(2005).
\bibitem{uni} H. Heiselberg Phys. Rev. A 63, 043606 (2001); K. M. O'Hara, S. L.
Hemmer, M. E. Gehm, S. R. Granade, J. E. Thomas, Science 298, 2179
(2002); G. M. Bruun, Phys. Rev. A 70, 053602 (2004); Tin-Lun Ho,
Phys. Rev. Lett. 92, 090402 (2004); J. Carlson, S.-Y. Chang, V. R.
Pandharipande, and K. E. Schmidt, Phys. Rev. Lett. 91, 050401 (2003)
\bibitem{vortexHe} S.C. Whitmore and W. Zimmermann, Jr., Phys. Rev. 166, 181
(1968).
\bibitem{vortexMIT} M. Zwierlein et al., Nature 435, 1047 (2005).
\bibitem{ho}  H. Zhai and T. -L. Ho, Phy. Rev. Lett., 97, 180414
(2006); M. Y. Veillette, D. E. Sheehy, L. Radzihovsky, and V.
Gurarie, Phys. Rev. Lett. 97, 250401 (2006); G. Moller and N. R.
Cooper, Phys. Rev. Lett. 99, 190409 (2007).
\bibitem{fqhe} N. R. Cooper, N. K. Wilkin, and J. M. Gunn, Phys. Rev.
Lett. 87, 120405 (2001); B. paredes, P. Zoller, and J. I. Cirac,
Sol. S. Com. 127, 155 (2003).
\bibitem{quarticBose} E. Lundh, Phys. Rev. A 65, 043604 (2002); K. Kasamatsu, M. Tsubota, and M. Ueda, Phys. Rev. A 66, 053606
(2002); G. Kavoulakis and G. Baym, New J. Phys. 5, 51.1 (2003); E.
Lundh, A. Collin, and K. A. Suominen, Phys. Rev. Lett. 92, 070401
(2004); T. K. Ghosh, Phys. Rev. A 69, 043606 (2004); T. K. Ghosh,
Eur. Phys. J. D 31 101 (2004); G. M. Kavoulakis, A. D. Jackson, and
Gordon Baym, Phys. Rev. A 70, 043603 (2004); Ionut Danaila, Phys.
Rev. A 72, 013605 (2005); A. Collin, Phys. Rev. A 73, 013611 (2006);
S. Bargi, G. M. Kavoulakis, and S. M. Reimann, Phys. Rev. A 73,
033613 (2006); Michiel Snoek and H. T. Stoof, Phys. Rev. A 74,
033615 (2006); S. Gautam, D. Angom, Eur. Phys. J. D 46, 151–155
(2008).
\bibitem{modesEX1} J. Kinast, A.
Turlapov, and J. E. Thomas, Phys. Rev. A 70, 051401(R) (2004); J.
Kinast, A. Turlapov, and J. E. Thomas, Phys. Rev. Lett. 94, 170404
(2005); M. Bartenstein, A. Altmeyer, S. Riedl, S. Jochim, C. Chin,
J. H. Denschlag, R. Grimm, Phys. Rev. Lett. 92, 203201 (2004).

\bibitem{modesEX2} J. Kinast, S. L. Hemmer, M. E. Gehm, A. Turlapov, and
J. E. Thomas, Phys. Rev. Lett. 92, 150402 (2004).

\bibitem{modesTH} Theja N. De Silva and Erich J. Mueller, Phys. Rev. A
72, 063614 (2005); H. Heiselberg, Phys. Rev. Lett. 93, 040402
(2004); S. Stringari, Europhys. Lett. 65, 749 (2004); H. Hu, A.
Minguzzi, X. J. Liu, and M. P. Tosi, Phys. Rev. Lett. 93, 190403
(2004); Y. E. Kim and A. L. Zubarev Phys. Rev. A 70, 033612 (2004);
N. Manini and L. Salasnich, Phys. Rev. A 71, 033625 (2005); G. E.
Astrakharchik, R. Combescot, X. Leyronas and, S. Stringari, Phys.
Rev. Lett. 95, 030404 (2005); A. Bulgac and G. F. Bertsch, Phys.
Rev. Lett. 94, 070401 (2005); M. Manini and L. Salasnich, Phys. Rev.
A 71, 033625 (2005); Y.Ohashi and A. Griffin, e-print
cond-mat/0503641.
\bibitem{vorticity} R. P. Feynman, edited by C. J. Gorter \emph{Progress in Low Temperature
Physics}, North- Holland, Amsterdam, 1955).
\bibitem{manini} N. Manini and L. Salasnich, Phys. Rev. A 71, 033625
(2005).
\bibitem{gosh} T. K. Gosh and K. Machida, Phy. Rev. A 73, 025601 (2006).
\bibitem{ant} M. Antezza, M. Cozzini, and S. stringari, Phys. Rev. A 75, 053609 (2007)
\bibitem{astra} G. E. Astrakharchik, J. Boronat, J. Casulleras, and S.
Giorgini, Phys. Rev. Lett. 93, 200404 (2004).
\end{references}
\end{document}